 \documentclass[aps,twocolumn,pra,superscriptaddress,amsmath,showpacs,tightenlines]{revtex4}

\usepackage{epsfig,graphicx,times}
\usepackage{amstext}
\usepackage{amsmath}            
\usepackage{amssymb}            
\usepackage{graphicx}           
\usepackage{latexsym}
\usepackage{bm}

\def\<{\langle}
\def\>{\rangle}

\def\extra#1{}

\begin{document}
\title{Qubit-induced phonon blockade as a signature of quantum behavior
in nanomechanical resonators}

\author{Yu-xi Liu} \affiliation{Advanced Science Institute, RIKEN, Wako-shi, Saitama
351-0198, Japan} \affiliation{Institute of Microelectronics,
Tsinghua University, Beijing 100084, China} \affiliation{Tsinghua
National Laboratory for Information Science and Technology
(TNList), Tsinghua University, Beijing 100084, China}

\author{Adam Miranowicz}
\affiliation{Advanced Science Institute, RIKEN, Wako-shi, Saitama
351-0198, Japan} \affiliation{Faculty of Physics, Adam Mickiewicz
University, 61-614 Pozna\'n, Poland}

\author{Y. B. Gao}
\affiliation{College of Applied Science, Beijing University of
Technology, Beijing, 100124, China}

\author{Ji\v r\'\i\ Bajer}
\affiliation{Department of Optics, Palack\'{y} University, 772~00
Olomouc, Czech Republic}

\author{C. P. Sun}
\affiliation{Institute of Theoretical Physics, The Chinese Academy
of Sciences, Beijing, 100080, China}

\author{Franco Nori}
\affiliation{Advanced Science Institute, RIKEN, Wako-shi, Saitama
351-0198, Japan} \affiliation{Physics Department, The University
of Michigan, Ann Arbor, Michigan 48109-1040, USA}

\date{\today}

\begin{abstract}
The observation of quantized nanomechanical oscillations by
detecting femtometer-scale displacements is a significant
challenge for experimentalists. We propose that phonon blockade
can serve as a signature of quantum behavior in nanomechanical
resonators. In analogy to photon blockade and Coulomb blockade for
electrons, the main idea for phonon blockade is that the second
phonon cannot be excited when there is one phonon in the nonlinear
oscillator. To realize phonon blockade, a superconducting quantum
two-level system is coupled to the nanomechanical resonator and is
used to induce the phonon self-interaction. Using Monte Carlo
simulations, the dynamics of the induced nonlinear oscillator is
studied via the Cahill-Glauber $s$-parametrized quasiprobability
distributions. We show how the oscillation of the resonator can
occur in the quantum regime and demonstrate how the phonon
blockade can be observed with currently accessible experimental
parameters.



\end{abstract}
\maketitle \pagenumbering{arabic}


\section{Introduction}

Many efforts (e.g., see
Refs.~\cite{Huang03,Knobel03,Blencowe04,LaHaye04} and
reviews~\cite{Blencowe04review,Schwab05review,Ekinci05review})
have been made to explore quantum effects in nanomechanical
resonators (NAMRs) and optomechanical systems (e.g., in
Refs.~\cite{add1,add2,add3,add4,add5,add6} and the
review~\cite{vahala08review}). Reaching the quantum limit of NAMRs
would have important applications in, e.g., small mass or
weak-force detections~\cite{Caves,Bocko96,Buks06}, quantum
measurements~\cite{Braginsky92}, and quantum-information
processing. Only recently the quantum limit in NAMRs has been
reached experimentally~\cite{OConnell}.

Quantum or classical behavior of a NAMR oscillation depends on its
environment, which induces the decoherence and dissipation of the
NAMR states.  In principle, if the NAMR is cooled to very low
temperatures (in the mK-range) and has sufficiently high
oscillation frequencies (in the GHz-range), then its oscillation
can approach the quantum limit. In other words, if the energies of
the NAMR quanta, which are referred to as {\em
phonons}~\cite{Cleland-book}, are larger than (or at least
comparable to) the thermal energy, then the mechanical oscillation
can be regarded quantum. When the NAMR can beat the thermal energy
and approach the quantum regime, measurements on quantum
oscillation of the NAMRs are still very challenging. One
encounters: (i) fundamental problems as measurements are usually
performed by the position detection,  the quantum uncertainty due
to the zero-point fluctuation will limit the measurement accuracy;
(ii) practical problems as, for a beam oscillating with frequency
in the gigahertz range, the typical displacement for this
oscillation is on the order of a femtometer. Detecting so tiny
displacement is a difficult task for current experimental
techniques.

Various signatures and applications of quantum behavior (or
nonclassicality) in nanomechanical resonators have been studied.
Examples include: generation of quantum
entanglement~\cite{Cleland04,Armour02,Tian05}, generation of
squeezed states~\cite{Hu96,Wang04,Rabl04,Xue07}, Fock
states~\cite{Santamore04,Buks08}, Schr\"odinger cat
states~\cite{Semiao09}, and other nonclassical
states~\cite{Tian04,Jacobs07}, prediction of
classical-like~\cite{Gronbech05} and quantum~\cite{Shevchenko08}
Rabi oscillations, transport measurements~\cite{Lambert}, quantum
nondemolition
measurements~\cite{Braginsky92,Irish03,Santamore04,Buks08,Gong08},
quantum tunneling~\cite{Savelev06}, proposal of quantum
metrology~\cite{Woolley08} and of quantum decoherence
engineering~\cite{Wang04}.

The problem of how to perform quantum measurements on a system
containing a NAMR plays a fundamental role in reaching the quantum
limit of the NAMR and testing its nonclassical behavior. Quantum
measurements are usually done by coupling an external probe
(detector) to the NAMR (see,
e.g.,~\cite{Knobel03,LaHaye04,Naik06,Wei06,Jacobs07,Srinivasan07,Regal08,Lambert08}
and references therein).

Our approach for detecting quantum oscillations of NAMRs is based
on: (i) recent theoretical proposals (e.g., Refs.~\cite{Gao}) to
perform quantum measurement on NAMR without using an external
probe and (ii) experimental demonstrations (e.g.,
Refs.~\cite{ntt,nec}) on the couplings between superconducting
quantum devices and the NAMRs. Instead of directly detecting a
tiny displacement, we propose to indirectly observe quantum
oscillations of the NAMR via {\em phonon blockade}, which is a
purely quantum phenomenon.

We assume that the phonon decay rate is much smaller than the
phonon self-interaction strength. In such a case, we show that
when the oscillations of the NAMR are in the nonclassical regime,
the phonon excitation can be blockaded. In analogy to the photon
(e.g., see Refs.~\cite{Imamoglu,Leonski94}) and Coulomb (e.g., see
Ref.~\cite{Coulomb}) blockades, the main idea for the phonon
blockade is that the {\em second phonon cannot be excited when
there is one phonon in the nonlinear oscillator.} Therefore, by
analyzing correlation spectra for the electromotive force
generated between two ends of the NAMR, the phonon blockade can be
distinguished from excitations of two or more phonons.

\begin{figure}
\includegraphics[bb=70 560 487 766, width=8.7 cm, clip]{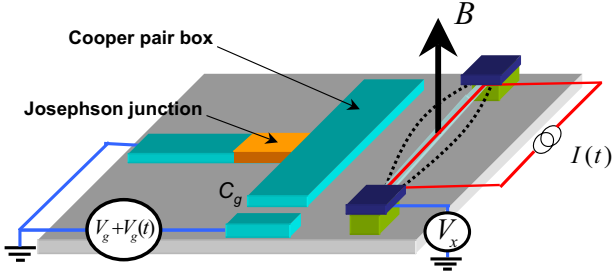}
\caption[]{(Color online) Schematic diagram for the coupling
between a nanomechanical resonator (light blue bar on the right
side) and superconducting charge qubit (left side). Two black
dashed curves on the right show that the resonator is oscillating.
The static magnetic field $B$ (presented by the black
upward-pointing arrow) and the alternating current $I(t)$ (shown
by the red loop on the right) are used for the motion detection of
the NAMR.}\label{fig1}
\end{figure}

An important ingredient for the realization of the phonon blockade
is strong phonon {\it self-interaction}. To obtain such {\it
nonlinear} phonon-phonon interaction, the NAMR is assumed to be
coupled to a superconducting two-level system, which can be either
charge, flux, or phase qubit circuits~\cite{you1,you2,you3,you4}.
By choosing appropriate parameters of two-level systems, a
nonlinear phonon interaction can be induced. The interactions
between each of these qubits~\cite{you1,you2,you3,you4} with NAMRs
are very similar, e.g., the coupling constants are of the same
order and the frequencies of these qubits are in the same GHz
range. Therefore, in this paper, we only use charge qubits as an
example to demonstrate our approach. However, this approach can
also be applied to demonstrate the oscillation of the NAMR in the
quantum regime, when the NAMR is coupled to other superconducting
qubits, e.g., phase or flux qubits.

The paper is organized as follows. In Sec.~II, we describe the
couplings between the superconducting qubits and the NAMR, and
then study how the qubit induces the phonon-phonon interaction. In
Sec.~III, we discuss how to characterize the quantum oscillation
by using the Cahill-Glauber $s$-parametrized quasiprobability
distributions for $s>0$, in contrast to the Wigner function (for
$s=0$). In Sec.~IV, the basic principle of the phonon blockade is
demonstrated, and we show that the phonon blockade can occur for
the different parameters. In Sec.~V, we study the measurement of
the phonon blockade by using the correlation spectrum of the
electromotive force between two ends of the NAMR. Finally, we
summary the main results of the paper in Sec.~VI.

\section{Qubit-induced phonon-phonon interaction}

Let us now focus on the coupling between a NAMR (with mass $m$ and
length $L$) and a superconducting charge qubit (with Josephson
energy $E_{J}$ and junction capacitance $C_{J}$). As schematically
shown in Fig.~\ref{fig1}, a direct-current (d.c.) voltage $V_{g}$,
and an a.c.~voltage $V_{g}(t)=V_{0}\cos(\omega_{1}t)$ are applied
to the charge qubit (or Cooper pair box) through the gate
capacitor $C_{g}$. The NAMR is coupled to the charge qubit by
applying a static voltage $V_{x}$ through the capacitor $C(x)$
which depends on the displacement $x$ of the NAMR around its
equilibrium position. A weak detecting current
$I(t)=I_{0}\cos(\omega_{2}t)$ is applied to the NAMR, with its
long axis perpendicular to the static magnetic field $B$.

In the rotating wave approximation and neglecting two-phonon terms,
the Hamiltonian $H=H_{0}+H_{\rm d}$ of the interaction system
between the charge qubit and the NAMR can be described
by~\cite{Gao}:
\begin{eqnarray}
H^{(0)}&=& \frac12 \hbar\omega_{0}\sigma_z+\hbar\omega
a^{\dagger}a
+\hbar g (a \sigma_{+}+a^{\dagger}\sigma_{-})\nonumber\\
&&\,+\hbar \Omega
\left(\sigma_{+}e^{-i\omega_{1}t}+\sigma_{-}e^{i\omega_{1}t}\right)\,,\label{eq:1}\\
H^{({\rm d})}&=&\hbar\epsilon \left(a^{\dagger}e^{-i\omega_{2}t}+a
e^{i\omega_{2}t}\right)\,. \label{eq:2}
\end{eqnarray}
Here, the frequency shift of the NAMR, due to its coupling to the
charge qubit, has been neglected because it just renormalizes the
NAMR frequency and will not affect the calculations below. This
frequency shift is determined~\cite{Tian04} by the qubit-NAMR
distance $l$, the charging energy $E_{c}=e^2/2(C_{J}+C_{g}+C)$,
the mass $m$ and the frequency $\omega$ of the NAMR. It should be
noted that below we consider the large detuning between the qubit
and the NAMR, i.e., $(\omega_{0}-\omega)$ is several times larger
(but not much larger) than the coupling constant $g$; thus, the
rotating wave approximation can be applied. The effect of the
counter-rotating terms on the results can also calculated in the
large detuning case~\cite{hanggi}. However, here we have neglected
this effect because it only produces a small frequency shift and
two-photon processes. The charge qubit, described by the spin
operator $\sigma_{z}=|e\rangle\langle e|-|g\rangle\langle g|$, is
assumed to be near the optimal point, i.e., $(C_{g}V_{g}+C
V_{x})/2e\approx 0.5$ with $C=C(x=0)$, and thus $\omega_{0}\approx
E_{J}/\hbar$. The qubit ground and excited states are denoted by
$|g\rangle$ and $|e\rangle$, respectively. The operator $a$
($a^{\dagger}$) denotes the  annihilation (creation) operator of
the NAMR with frequency $\omega$,  which can be written as
\begin{eqnarray}
a&=& \sqrt{\frac{m\omega}{2\hbar}}\left(x+\frac{i}{m\omega}p\right),\\
a^{\dagger}&=&
\sqrt{\frac{m\omega}{2\hbar}}\left(x-\frac{i}{m\omega}p\right)
\end{eqnarray}
with the momentum operator $p$ of the NAMR. The third term of
Eq.~(\ref{eq:1}) presents the NAMR-qubit interaction with the
strength
\begin{eqnarray}
  g=\frac{4E_{c}N_{x} X_{0}}{d}
\label{e1}
\end{eqnarray}
determined by the charging energy $E_{c}$, effective Copper pair
number $N_{x}=CV_{x}/2e$, the distance $d$ between the NAMR and
the superconducting qubit, and the NAMR amplitude
$X_{0}=\sqrt{\hbar/2m\omega}$ of zero-point motion. Also, $\Omega$
is the Rabi frequency of the qubit driven by the classical field
with frequency $\omega_{1}$. The parameter
\begin{eqnarray}
  \epsilon=-BI_{0}L X_{0}
\label{e2}
\end{eqnarray}
in Eq.~(\ref{eq:2}) describes the interaction strength between the
NAMR and an external weak probe a.c.~current with frequency
$\omega_{2}$. Hereafter, we assume that the resonant driving
condition for the qubit is satisfied, i.e., $\omega_1=\omega_0$.
For the coupling between a phase~\cite{Cleland04} (or flux
qubit~\cite{ntt}) and the NAMR, they also have the same form as
that given in Eqs.~(\ref{eq:1}) and (\ref{eq:2}) except all
parameters of the Hamiltonian should be specified to the concrete
systems. Thus, our discussions below can also be applied to those
systems.

The frequency of the NAMR is usually much lower than that of the
qubit. If the Rabi frequency $\Omega$ satisfies the condition
$\Omega \gg (g^2/\Delta)$ with the detuning
$\Delta=\omega_{0}-\omega$ between the frequencies of the NAMR and
the qubit, then in the rotating reference frame with
$V=\exp(-i\omega_{0}\sigma_{z}/2)$, Eq.~(\ref{eq:1}) is equivalent
to an effective Hamiltonian
\begin{equation}
H_{\rm eff}^{(0)}=\hbar\omega a^{\dagger}a+\hbar
\left[\frac{g^2}{\Delta} a^{\dagger}a -\kappa (a^{\dagger} a)^2
\right]\rho_{z}
\end{equation}
with the effective phonon self-interaction constant (nonlinearity
constant)
\begin{eqnarray}
  \kappa=\frac{g^4}{\Omega\Delta^2}\,.
\label{e3}
\end{eqnarray}
Here, $\rho_{z}=|+\rangle\langle +|-|-\rangle\langle -|$ with the
dressed qubit states $|\pm\rangle=(|g\rangle \pm
|e\rangle)/\sqrt{2}$. Therefore, if the dressed charge qubit, which
was theoretically proposed~\cite{liu06} and has been experimentally
realized~\cite{delsing1,delsing2}, is always in its ground state
$|-\rangle$, the effective Hamiltonian for the driven NAMR is
\begin{equation}
H_{\rm eff}=\hbar \left(\omega-\frac{g^2}{\Delta}
\right)a^{\dagger}a +\hbar\kappa (a^{\dagger}a)^2+\hbar\epsilon
(a^{\dagger}{\rm e}^{-i\omega_{2} t}+a{\rm e}^{i\omega_{2}
t}).\label{eq:4-1}
\end{equation}

The nonlinear Hamiltonian of the driven NAMR in Eq.~(\ref{eq:4-1})
can also be directly obtained when the driving field is strong;
however, here we only consider a weak probe current. Thus, the
coupling of the NAMR to a controllable superconducting two-level
is necessary for inducing phonon-phonon interactions.

\begin{figure}

\centerline{\epsfxsize=8cm\epsfbox{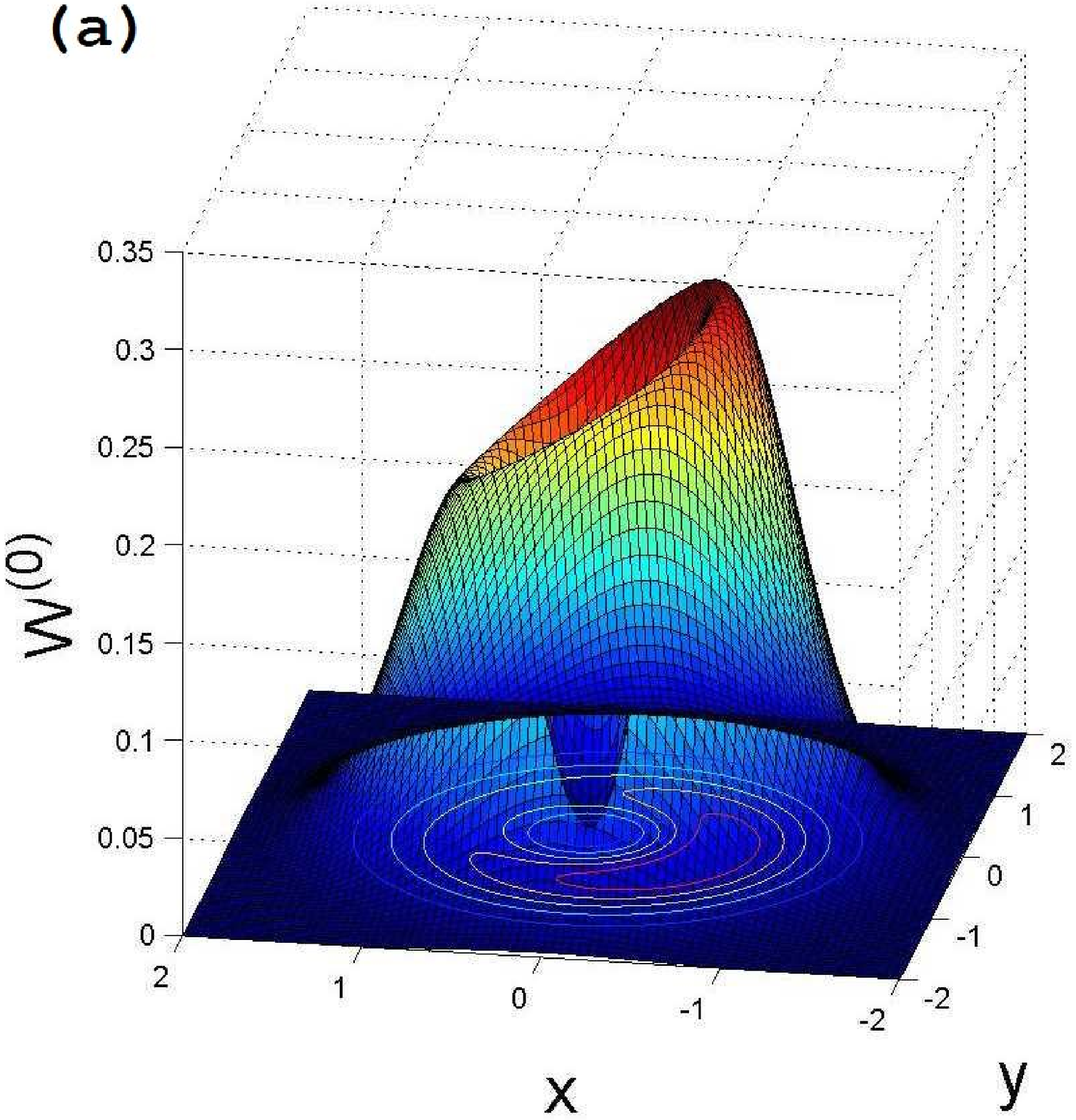}}

\centerline{\epsfxsize=8cm\epsfbox{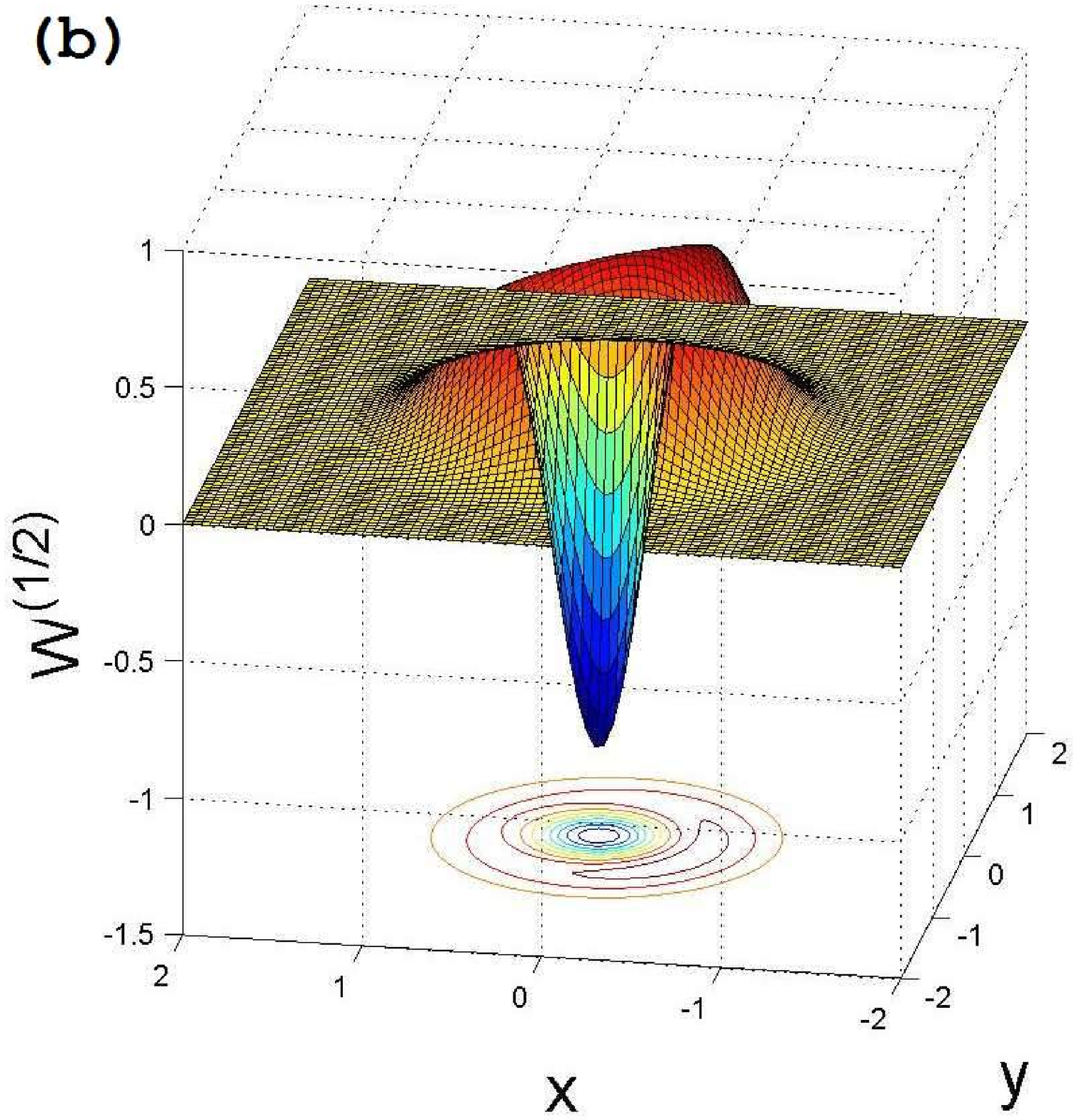}}

\caption{(Color online) Quasidistribution functions for the NAMR
steady state obtained by solving master equation (\ref{N03}) for
$\epsilon=3\gamma, \kappa=30\gamma$, and $\bar{n}=0.01$ with
$\gamma$ as units: (a) Wigner function $W^{(0)}(x,y)$, which is
non-negative in the whole phase space, and (b) 1/2-parametrized
quasi-probability distribution (QPD) $W^{(1/2)}(x,y)$, which is
negative for $\alpha=x+iy$ close to zero indicating
nonclassicality of the NAMR state. The figures show the bottom of
the functions.}\label{fig2}
\end{figure}

\section{Quantum behavior described by quasiprobability distributions}

Decoherence imposes strict conditions to observe quantum behavior
in a NAMR. To demonstrate effects of the environmental on the
NAMR, let us now assume that the NAMR is coupled to the thermal
reservoir. Under the Markov approximation, the evolution of the
reduced density operator $\rho$ for the NAMR can be described by
the master equation~\cite{Carmichael}:
\begin{eqnarray}
\frac{\partial }{\partial t}{\rho} &=&-\frac{i}{\hbar}[H_{\rm
eff},{\rho} ] +\frac{\gamma }{2}\bar{n}(2 a^{\dag}\rho a-a
a^{\dag}\rho-\rho a a^{\dag})
\nonumber \\
&& +\frac{\gamma }{2}(\bar{n}+1)(2a\rho
a^{\dag}-a^{\dag}a\rho-\rho a^{\dag}a). \label{N03}
\end{eqnarray}
In Eq.~(\ref{N03}), $\gamma $ is the damping rate and
$\bar{n}=\{\exp [\hbar \omega /(k_{B}T)]-1\}^{-1}$ is the mean
number of thermal phonons,  where $k_{B}$ is the Boltzmann
constant, and $T$ is the reservoir temperature at thermal
equilibrium. Eq.~(\ref{N03}) can be solved, e.g., by applying the
Monte Carlo wave function
simulation~\cite{Carmichael,Dalibard,Tan} and introducing the
collapse operators
\begin{eqnarray}
C_1=\sqrt{\gamma(\bar n+1)}a\,,\quad C_2=\sqrt{\gamma\bar
n}a^\dagger\,. \label{N03a}
\end{eqnarray}

We now study the steady-state solution, which is independent of the
initial states. For the system without a drive, the time evolution
ends in a state without phonons (vacuum state) at zero temperature.
While for a driven system, the asymptotic state is neither the
vacuum nor a pure state even at zero temperature, and can have
intriguing noise properties.

A state is considered to be {\em nonclassical} if its
Glauber-Sudarshan $P$ function cannot be interpreted as a
probability density, i.e., it is negative or more singular than
Dirac's $\delta$ function. Due to such singularities, it is
usually hard to visualize it. To characterize the nonclassical
behavior of the NAMR states generated in our system, we consider
the Cahill-Glauber $s$-parametrized quasiprobability distribution
(QPD) functions~\cite{Cahill69}:
\begin{eqnarray}
  {\cal W}^{(s)}(\alpha) &=& \frac{1}{\pi} \,{\rm Tr}\,[ {\rho}\,
  {T}^{(s)}(\alpha)]\,,
\label{N09}
\end{eqnarray}
where
\begin{equation}
{T}^{(s)}(\alpha) = \frac1{\pi} \int \exp(\alpha\xi^*-\alpha^*\xi)
{D}^{(s)}(\xi) \,{\rm d}^2 \xi\,,
\end{equation}
and
\begin{equation}
{D}^{(s)}(\xi) = \exp\left(s\frac{|\xi|^2}{2}\right) {D}(\xi)\; ,
\end{equation}
with
\begin{equation}
{D}(\xi)=\exp\left(\xi a^\dagger-\xi^{*}a\right)\; ,
\end{equation}
being the displacement operator. The QPD is defined for $-1 \le
s\le 1$, which in special cases reduces to the $P$ function (for
$s=1$), Wigner function (for $s=0$), and Husimi $Q$ function (for
$s=-1$). QPDs contain the full information about states.

Let us analyze the differences between the 1/2-parametrized QPD and
Wigner functions under the resonant driving for the NAMR with
$\omega_{2}=\omega-(g^2/\Delta)$. As an example, in Fig.~\ref{fig2},
we plotted the steady-state Wigner function and 1/2-parametrized
QPD, which are the numerical solutions of the master equation for a
set of parameters: $\bar{n}=0.01$, $\epsilon=3\gamma$, and
$\kappa=30\gamma$ in units of $\gamma$. Fig.~\ref{fig2}(a) shows the
non-negative Wigner function of the steady state of the NAMR for
these parameters. It can also be shown analytically that the
steady-state Wigner function for this system is always non-negative.
However, the plot for the QPD function ${\cal W}^{(1/2)}(\alpha)$ in
Fig.~\ref{fig2}(b), with the same parameters as for
Fig.~\ref{fig2}(a), clearly shows negative values, corresponding to
a nonclassical state of the NAMR. Below, we will discuss how to
demonstrate this nonclassicality of the NAMR via the phonon
blockade.

The Wigner function for the NAMR steady state is non-negative in
the whole phase space. This is in contrast to the Wigner function
for various nonclassical states, including Fock states or finite
superpositions of coherent states (often referred to as
Schr\"odinger cat states) being negative in some areas of phase
space. It should be noted that there are other well-known
nonclassical states, including squeezed states, for which the
Wigner function is non-negative as for the NAMR steady state. In
general, the non-positivity of the Wigner function is a necessary
but not a sufficient condition for the non-classicality. The
complete characterization of the non-classicality (the ``if and
only if'' condition) is based on the positivity of the
$P$-function. Unfortunately, this function is usually too singular
to be presented graphically. The larger parameter $s$ the more
nonclassical states are described by the negative $s$-parametrized
QPD. In our case, to demonstrate the nonclassically of the NAMR
steady state, it was enough to calculate the $s$-parametrized QPD
for $s=1/2$ but not for $s=0$.

\begin{figure}
\centerline{\epsfxsize=8.2cm\epsfbox{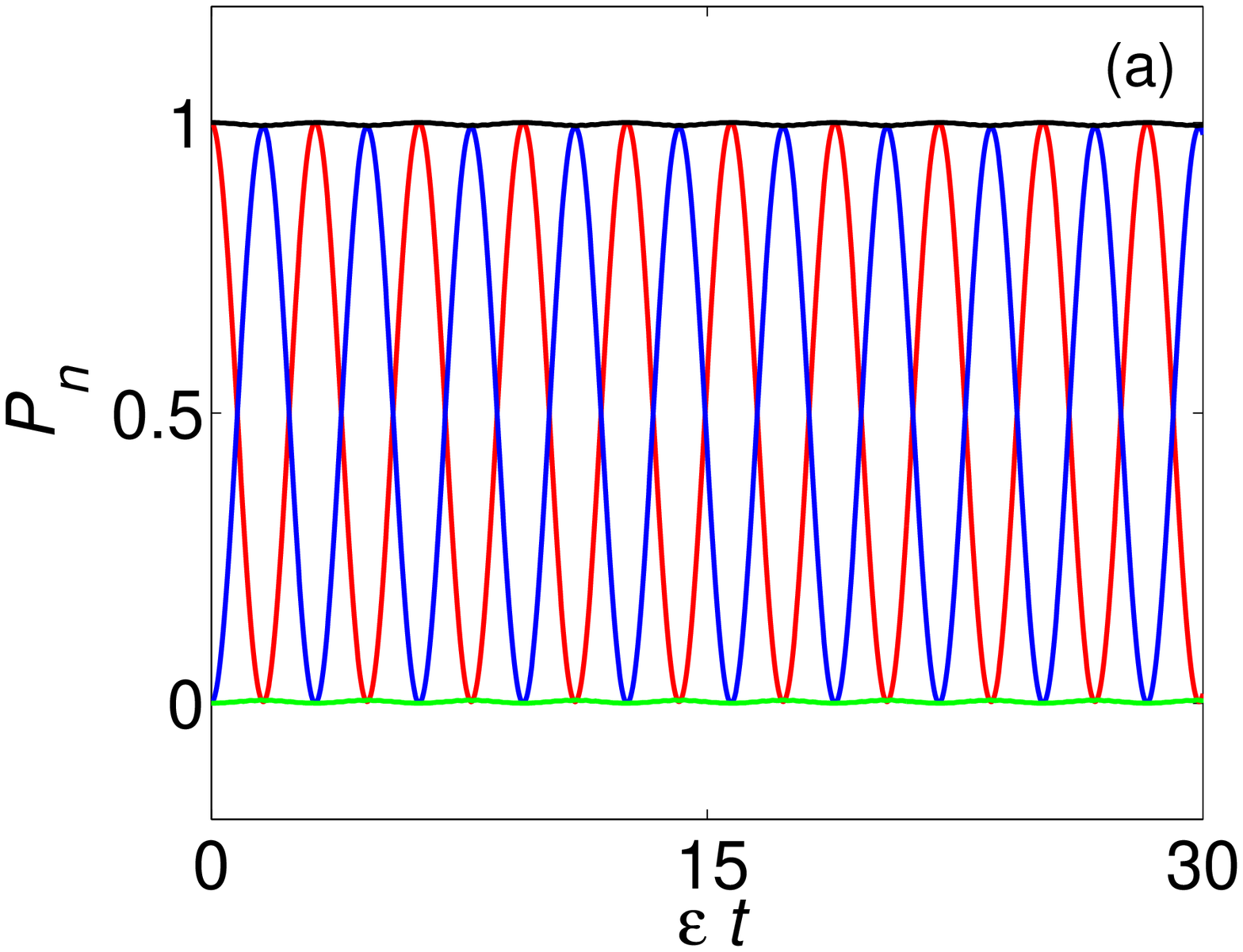 }}
\centerline{\hspace*{7mm}\epsfxsize=8.9cm\epsfbox{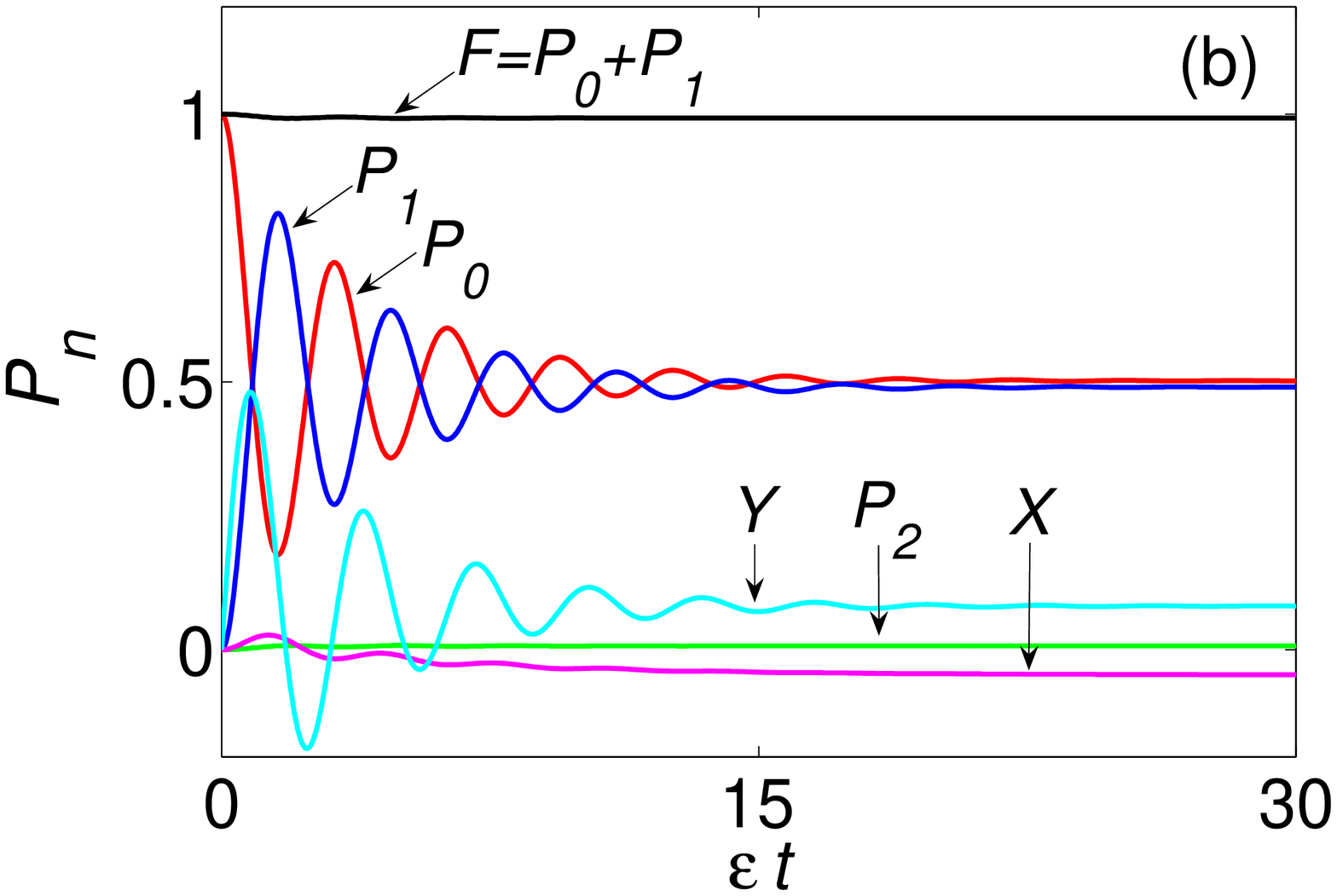}}

\vspace*{-1mm} \caption{(Color online) Probabilities $P_n=\langle
n|\rho(t)|n\rangle$ of measuring $n$ phonons as a function of
rescaled time, $\epsilon t$, assuming $\kappa=10\epsilon$ and: (a)
no dissipation ($\gamma=0$) and (b) including dissipation with the
same parameters as in Fig.~\ref{fig2}: $P_0$ (red curves), $P_1$
(blue), and $P_2$ (green). $F=P_0+P_1$ (thick black) describes the
fidelity of the phonon blockade. Additionally, the coherences
$X={\rm Re}\<0|\rho|1 \>$ (magenta curves) and $Y={\rm
Im}\<0|\rho|1\>$ (cyan) show that the steady states partially
preserve coherence. }\label{fig3}
\end{figure}

\section{Phonon blockade}

We now consider the case when the {\it phonon self-interaction}
strength $\kappa$ is much larger than the phonon decay rate
$\gamma$. When the oscillation of the NAMR is in the quantum regime,
the phonon transmission {\it can be blockaded} in analogy to the
single-photon blockade in a cavity~\cite{Imamoglu,Leonski94}. This
is because the existence of the second phonon requires an additional
energy $\hbar\kappa$.

To demonstrate the phonon blockade, let us rewrite
Eq.~(\ref{eq:4-1}) as
\begin{equation}
H_{\rm eff}=\hbar \bar \omega a^{\dagger}a +\hbar\kappa
a^{\dagger}a(a^{\dagger}a-1)+\hbar\epsilon (a^{\dagger}{\rm
e}^{-i\omega_{2} t}+a{\rm e}^{i\omega_{2} t})\label{eq:4}
\end{equation}
with a renormalized frequency
\begin{eqnarray}
  \bar \omega=\omega +\kappa-\frac{g^2}{\Delta}.
\label{eq:4a}
\end{eqnarray}
In the rotating reference frame for $V^{\prime}=\exp(-i\omega_{2}
a^{\dagger}a \,t)$ with $\omega_{2}=\bar\omega$, the Hamiltonian
in Eq.~(\ref{eq:4}) becomes
\begin{equation}
H_{\rm eff}=\hbar\kappa a^{\dagger}a(a^{\dagger}a-1)+\hbar\epsilon
(a^{\dagger}+a).\label{eq:4-2}
\end{equation}
It is now easy to see that the two states $|0\rangle$ and
$|1\rangle$ with zero eigenvalues are degenerate in the first term
$\kappa a^{\dagger}a(a^{\dagger}a-1)$ of Eq.~(\ref{eq:4-2}). This
degeneracy plays a crucial role in the phonon blockade. Indeed, if
we assume that the interaction strength $\epsilon$ is much smaller
than the nonlinearity constant $\kappa$ (i.e., $\epsilon
\ll\kappa$), then the phonon eigenstates of the Hamiltonian in
Eq.~(\ref{eq:4-2}) can become a superposition of only two states,
$|0\rangle$ and $|1\rangle$, in the lowest-order approximation of
the expansion in the strength $\epsilon$.

We now study the solution of the Hamiltonian in Eq.~(\ref{eq:4})
under the assumption of a weak driving current, i.e.,
$\epsilon\ll\kappa$. Using standard perturbation theory, the state
governed by the time-dependent periodic Hamiltonian in
Eq.~(\ref{eq:4}) with the initial condition
$|\psi(t=0)\rangle=|0\rangle$ can be obtained by introducing the
auxiliary operator
\begin{equation}
H_{F}=H_{\rm eff}-i\frac{\partial}{\partial t}
\end{equation}
based on the Floquet theory (e.g., see Ref.~\cite{Peskin}). The
solution can be approximately given as
\begin{equation}
  |\psi(t)\rangle = \cos(\epsilon t)|0\rangle
  -i \sin(\epsilon t)|1\rangle +{\cal O} (\epsilon^2).
  \label{N10}
\end{equation}
The solution~in Eq.~(\ref{N10}) shows that the number of phonons
varies between $0$ and $1$ if all terms proportional to
$\epsilon^2$ are neglected. In this small $\epsilon$ limit, the
Floquet solution~(\ref{N10}) explicitly demonstrates the {\em
phonon} blockade in analogy to the photon blockade~\cite{Imamoglu}
or the Coulomb blockade~\cite{Coulomb}, i.e., there is only
one-phonon excitation and the excitation with more than one phonon
is negligibly small. The photon blockade is also referred to as
the optical state truncation~\cite{Leonski94,Leonski01}.

The phonon-blockaded state is nonclassical as it is a
superposition of a finite number (practically two) of Fock states.
Only (some) superpositions of an infinite number of Fock states
can be considered classical.

The time-dependent probabilities $P_n=\langle n|\rho(t)|n\rangle$
of measuring the $n$-phonon state with and without dissipation are
numerically simulated using the Monte Carlo method. In the ideal
non-dissipative case, as shown in Fig.~\ref{fig3}(a), the sum of
the probabilities $P_{0}$ and $P_{1}$ with phonon numbers $0$ and
$1$ is almost one, which means that phonon blockade occurs. For
the dissipative case, Fig.~\ref{fig3}(b) shows the time evolutions
of the elements $\langle m |\rho(t)|n \rangle $ (with
$m,\,n=0,\,1$) for the same parameters as those in
Fig.~\ref{fig2}. The amplitudes of $P_{0}$ and $P_{1}$ exhibit
decaying oscillations; however, their sum is still near one and
thus the sum of other probabilities $P_{n}$ with $n> 1$ is near
zero. Therefore, the phonon blockade occurs even in the long-time
limit (e.g., steady state).  The non-zero off-diagonal element
$\langle 0|\rho(t)|1 \rangle$ shown in Fig.~\ref{fig3}(b) in the
steady-state means that the NAMR is in the nonclassical state,
which is also consistent with the steady-state plot of the QPD in
Fig.~\ref{fig2}(b). Thus, we see that the non-negative Wigner
function does not directly indicate that the state is
nonclassical.

\begin{figure}
\centerline{\hspace*{7mm}\epsfxsize=9.2cm\epsfbox{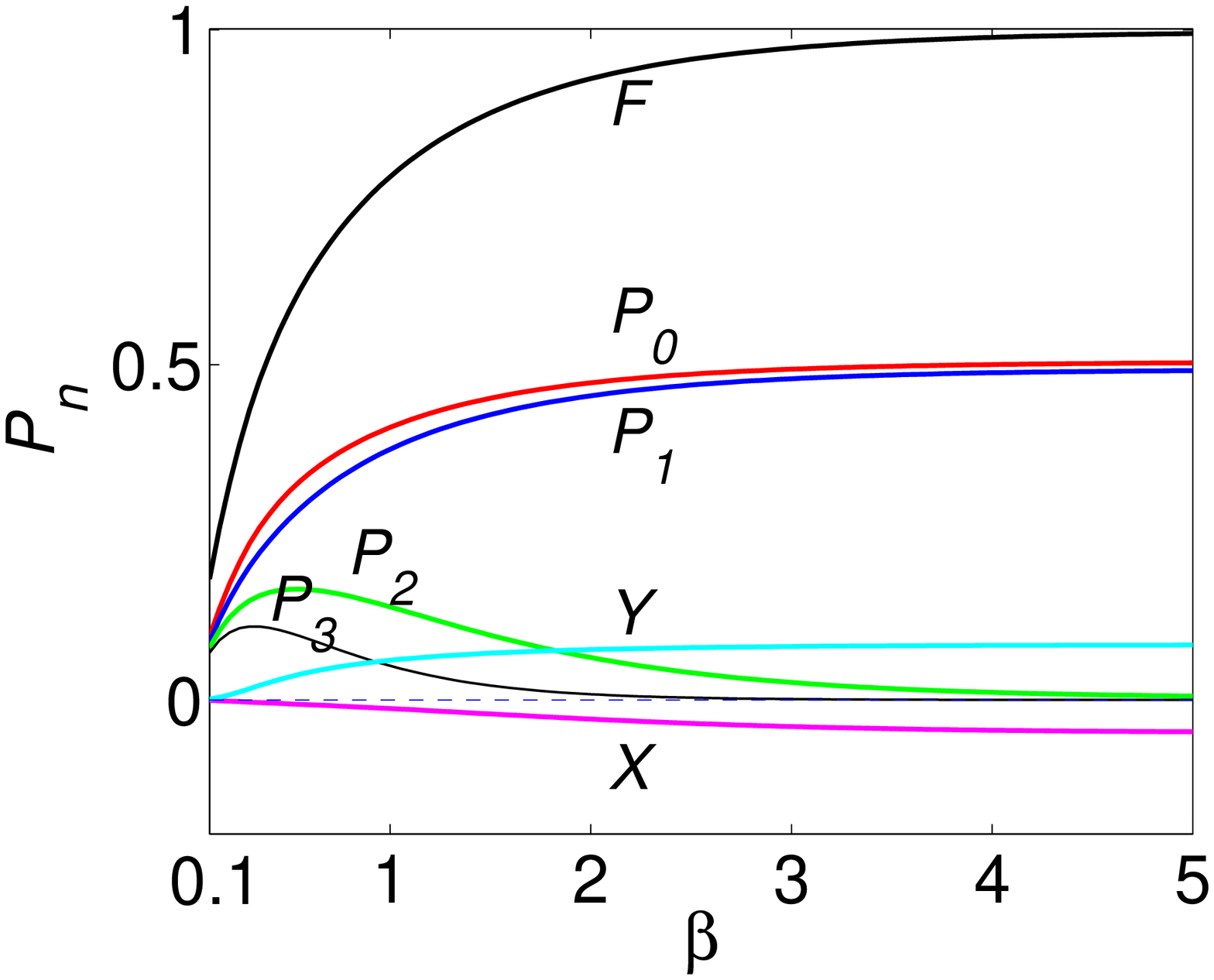}}
\centerline{\epsfxsize=8.5cm\epsfbox{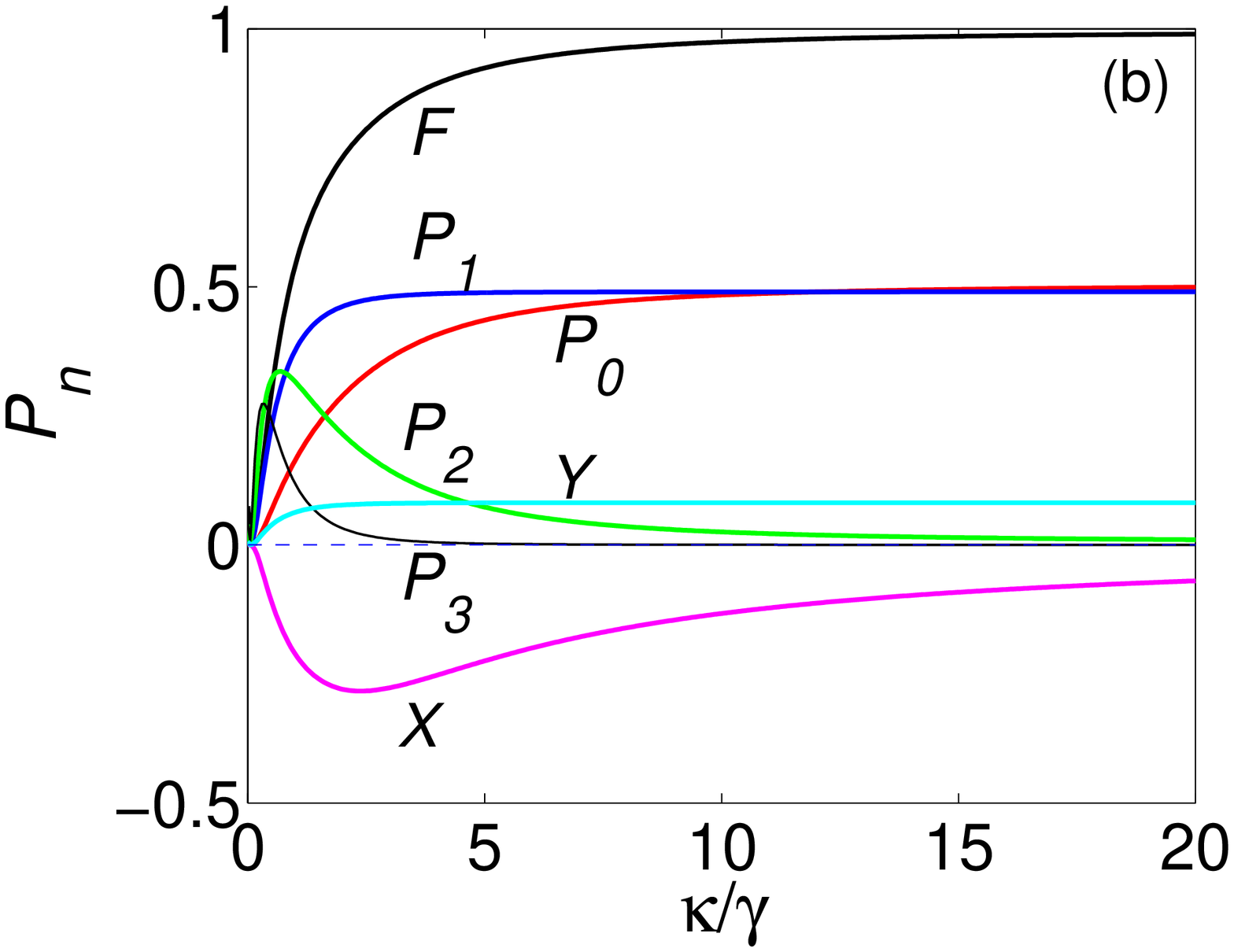}}

\vspace*{-1mm} \caption{(Color online) Probabilities $P_n$,
fidelity $F$, and coherences $X$ and $Y$ for steady states as a
function of (a) $\beta=\hbar \omega /(k_{B}T)$, assuming
$\kappa=10\epsilon$, and of (b) $\kappa/\gamma$, assuming
$\bar{n}=0.01$, which corresponds to $\beta\approx 4.6$. In both
(a) and (b), we set $\gamma=1$. The other parameters are the same
as in Fig.~\ref{fig3}.}\label{fig4}
\end{figure}
To study how the environmental temperature $T$ affects the phonon
blockade, the probability distributions $P_{n}=\langle
n|\rho(t)|n\rangle$ (for $n=0,\,1,\,2,\,3$) are plotted via the
rescaled inverted temperature $\hbar\omega/(k_{B}T)$ in
Fig.~\ref{fig4}(a). It clearly shows that the phonon blockade
cannot be achieved when the thermal energy is much larger than
that of the NAMR.  The $\kappa$-dependent matrix elements $\langle
m|\rho(t)|n \rangle$ are plotted in Fig.~\ref{fig4}(b), which
shows that the larger nonlinearity parameter $\kappa$ corresponds
to a more effective phonon blockade. However, to observe the
phonon blockade, it is enough to make $\kappa$ larger than a
certain value. For instance, if the ratio between $\kappa/\gamma$
is larger than $10$, then the sum of the probabilities $P_{0}$ and
$P_{1}$ is more than $0.95$, and the phonon blockade should occur.

Let us make a few comments to clarify the relation between the
phonon blockade and nonclassicality in terms of the
$s$-parametrized QPDs: (i) If the $s$-parametrized QPD, for some
$s\in(-1,1]$ and for a given state, is negative in some region of
the phase space, then the state is nonclassical. (ii) Even if the
phonon blockade is not observed (for a given choice of parameters
$\epsilon$, $\kappa$, $\gamma$, and $\bar n$), the
$1/2$-parametrized QPD (or the QPD for any $s>-1$) can still be
nonpositive. (iii) Even if we choose the parameters such that the
$1/2$-parametrized QPD is positive, this does not imply that the
state is classical. (iv) Even if the phonon blockade does not
appear, the state can be nonclassical as described by the
nonpositive $P$-function (the QPD for $s=1$).

A good blockade of phonons can be observed for nonclassical states
only. However, a poor blockade of phonons does not imply that the
state is classical. Similarly to other quantum effects like
squeezing or antibunching: If a specific nonclassical effect is
not exhibited by a given state, it does not imply that the state
is classical.

We can choose the parameters $\epsilon$, $\kappa$, $\gamma$, and
$\bar n$ in order to observe a change (transition) from a
nonpositive $1/2$-parametrized QPD to a positive function.
However, this transition is not important in the context of
nonclassicality. For various $s>-1$, one could observe such
transitions for different parameters. Only the transition of the
$P$-function corresponds to a transition from quantum to classical
regime. As already mentioned, a good criterion of nonclassicality
should be based on the $P$-function, but it is usually too
singular to be presented graphically. Thus, we have chosen the QPD
for another value of $s\in(0,1)$. A nonclassicality criterion
based on the QPD for $s=1/2$ is more sensitive than that based on
the Wigner function (the QPD for $s=0$), but still it is not
sensitive enough in the general case, i.e., there are nonclassical
fields described by the positive $1/2$-parametrized QPD.

\section{Proposed measurements of the phonon blockade}

Let us now discuss how to measure the phonon blockade via the
magnetomotive technique, which is one of the basic methods to
detect the motion of NAMRs~\cite{small}. As shown in
Fig.~\ref{fig1}, the induced electromotive force $V$ between two
ends of the NAMR is~\cite{Gao,small}
\begin{equation}
V=BL\frac{p}{m}=iBL\sqrt{\frac{\hbar
\omega}{2m}}(a^{\dagger}-a)\,,
\end{equation}
which can be experimentally measured as discussed in
Ref.~\cite{small}. Here, $p$ is the momentum for the center of the
NAMR mass.  We analyze the power spectrum
\begin{equation}
S_{V}(\omega^{\prime})=\int_{-\infty}^{\infty}\langle
V(0)V(\tau)\rangle e^{-i\omega^{\prime}t} d t
\end{equation}
defined by the Fourier transform of the induced
electromotive-force two-time correlation function
\begin{equation}
\langle V(0)V(\tau)\rangle\equiv \lim_{t\rightarrow\infty}\langle
V(t)V(t+\tau)\rangle\,.
\end{equation}
This power spectrum can be measured effectively.

\begin{figure}
\epsfxsize=8.7cm\epsfbox{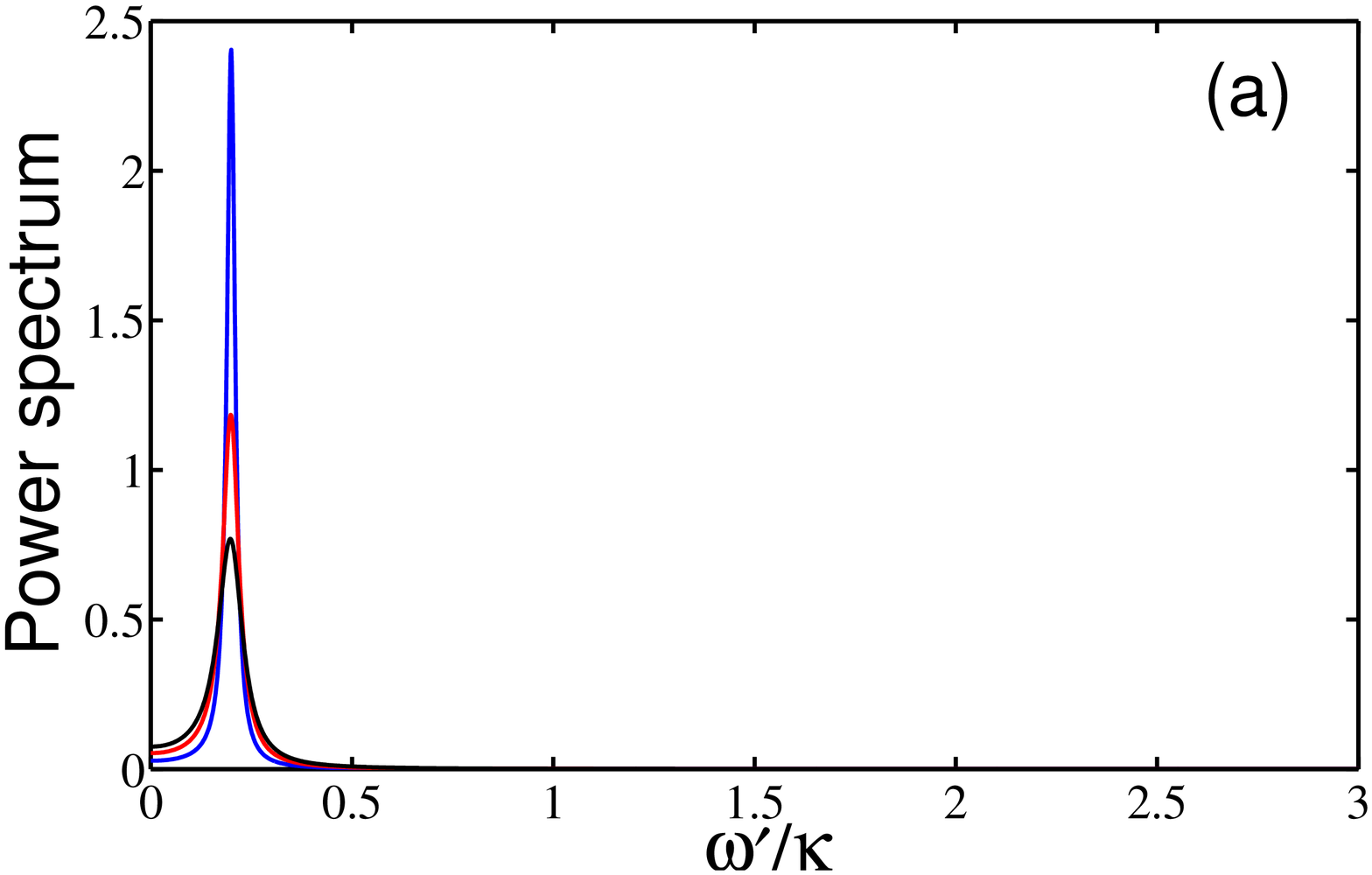}

\epsfxsize=8.7cm\epsfbox{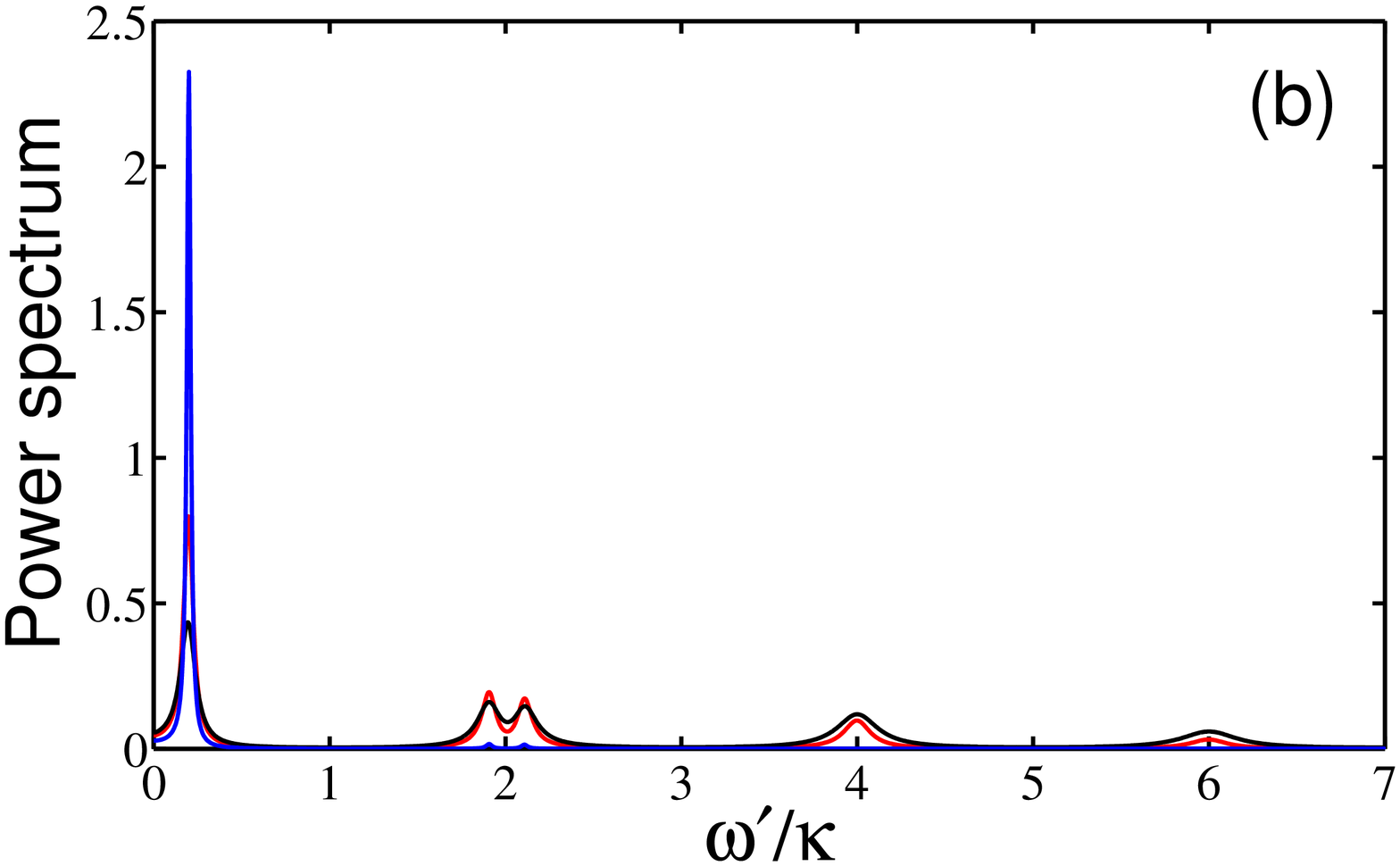}

\caption{(Color online) Power spectra $S_{V}(\omega^{\prime})$ for
$\kappa=30,$ $\epsilon=3$ and: (a) $\bar n=0$ with $\gamma=0.5$
(blue), 1 (red), 1.5 (black), and (b) $\gamma=0.5$ with $\bar
n=0.01$ (blue), 0.5 (red), 1 (black curves). Parameters $\kappa$,
$\epsilon$ and $\gamma$ are in units of $g$ on the order of MHz.}
\label{fig5}
\end{figure}

Power spectrum $S_{V}(\omega^{\prime})$ and the two-time
correlation function $\langle V(0)V(\tau)\rangle$ are plotted for
zero temperature with different decay rates $\gamma$ in
Fig.~\ref{fig5}(a), and for a given decay rate with different
temperatures $T$ (i.e., different thermal phonon number $\bar{n}$)
in Fig.~\ref{fig5}(b). We find that low dissipation and low
temperatures produce high spectral peaks, which enable an easier
observation of the phonon blockade. Thus, the environment (or some
background) will limit the power spectrum for observing the phonon
blockade. When $\kappa$ is negligible compared with the decay rate
$\gamma$, all spectral peaks disappear and there is no phonon
blockade. By other numerical calculations, we also find that a
large or giant nonlinearity $\kappa$ corresponds to sharp peaks
and, in this case, the phonon blockade is also easy to be
observed.

Assuming perfect phonon blockade, i.e., truncation to an exact
qubit state, one can analyze the whole evolution of our system
confined in two-dimensional Hilbert space. To some extend this
approximation can be applied in our model if the conditions
$\epsilon \ll \kappa $ and $\langle \bar{n}\rangle \approx 0$ are
satisfied. Then, we find that the corresponding power spectrum
should have at most three peaks at frequencies
\begin{eqnarray}
\omega'_{0}=0,\quad \omega'_{1,2}=\pm\frac{1}{4}
\sqrt{(8\epsilon)^{2}-\gamma^{2}
\left(1+2\bar{n}\right)^{2}}\approx \pm 2\epsilon. \label{ps1}
\end{eqnarray}
It is seen that these frequencies are independent of $\kappa$. A
peak at $\omega'_{0}=0$ does not appear for real $\epsilon$, which
is the case analyzed in the paper. Examples of such power spectra
for $\omega'>0$ are shown in Fig. 5(a) for $\bar{n}=0$ and in Fig.
5(b) for $\bar{n}=0.01$ (blue curve). In contrast, new peaks
appear in the spectrum in the case of not perfect phonon blockade.
This can be understood by analyzing a Hilbert space of dimension
$d>2$. For example, by analyzing the system evolution confined in
a three-dimensional Hilbert space, we find that the spectrum can
have at most seven peaks centered at
\begin{equation}
\omega ^{\prime }\approx 0,\pm 2\epsilon \left( 1-\delta \right)
,\pm \left[ 2\kappa \left( 1+6\delta\right) \pm \epsilon \left(
1-\delta\right) \right] , \label{ps2}
\end{equation}
where $\delta=\epsilon ^{2}/(8\kappa ^{2})$, which depend on
$\kappa$, contrary to Eq.~(\ref{ps1}). Frequencies in
Eq.~(\ref{ps2}) can be approximated as $\omega'\approx 0,\pm
2\epsilon,\pm (2\kappa\pm\epsilon)$. Thus, for $\omega'>0$, the
first peak occurs at $2\epsilon$, which corresponds approximately
to $\omega'_{1}$ given in Eq.~(\ref{ps1}). The second
characteristic double peak is at $2\kappa\pm\epsilon$, as seen in
Fig. 5(b) for $\bar{n}=0.5$ (red) and $\bar{n}=1$ (black curves).
Eq.~(\ref{ps2}) explains only the occurrence of the first three
peaks for $\omega'>0$ in Fig. 5(b). To explain the appearance of
the other two peaks at $\omega'\approx 4\kappa$ and $6\kappa$, one
should analyze the evolution of our system confined in (at least)
four-dimensional Hilbert space. Thus, these extra peaks are a
signature of a non-perfect single phonon blockade.

The spectra are not symmetric in frequency around zero,
$S_{V}(\omega^{\prime})\neq S_{V}(-\omega^{\prime})$.
Nevertheless, we depicted only the positive-frequency half of the
spectra in Fig. 5 to better compare the peaks for different values
of $\bar{n}$. We note that a double peak is observed at negative
frequencies $\omega'\approx - (2\kappa\pm\epsilon)$ even for the
cases shown in Fig. 5(a). This means that the contribution of
terms ${\cal O} (\epsilon^2)$ in Eq.~(\ref{N10}) is not negligible
for the parameters chosen in Fig. 5(a), and the spectrum for
$\omega'<0$ does not correspond to a (mathematically) perfect
single-phonon blockade.

In Fig. 5(b), the power spectra are plotted as a function of
$\omega'/\kappa$. There, it is seen that the position of the first
positive peak depends on the ratio $\epsilon/\kappa$ in agreement
with Eq.~(\ref{ps2}). The center of this peak is closer to zero
for smaller ratio $\epsilon/\kappa$. However, the position of the
center of the double peak (split peak) is approximately
independent of $\epsilon$ and $\kappa$ (assuming that
$\kappa\gg\epsilon$, so $\delta\approx 0$), which follows from
Eq.~(\ref{ps2}). Moreover, the splitting vanishes with increasing
$\gamma$. The smaller $\epsilon$ the smaller is $\gamma$ for which
the splitting vanishes.

In conclusion, the observation of extra peaks at frequencies
different from those in Eq.~(\ref{ps1}), show deterioration of the
single-phonon blockade. The higher are such peaks the worse is the
phonon blockade.

Note that the double peak at $\omega^{\prime}\approx
2\kappa\pm\epsilon$ was found assuming the output state to be in a
qutrit (three-dimensional) state. This double peak can, in
general, be predicted for a qudit state, i.e., $d$-dimensional
state for $2<d\ll\infty$. This corresponds to a phonon-truncation
up to state $|d-1\rangle$ and can be interpreted as a generalized
multi-phonon blockade. Any qudit states are nonclassical since
arbitrary finite superpositions of number states are nonclassical.
However, with increasing dimension $d$ of qudit states it becomes
more difficult to distinguish them from classical
infinite-dimensional states generated in our system. For this
reason, here we analyze the standard single-phonon blockade only.

We now discuss the experimental feasibility of our proposal. With
current experiments on coupling a superconducting
phase~\cite{Cleland04} (or charge~\cite{Tian04,Rabl04,Tian05} or
flux~\cite{ntt}) qubit and the NAMR, the coupling constants are on
the order of hundreds of MHz (e.g., $200$ MHz), the environmental
temperature can reach several tens of mili-Kelvin (e.g., $20$ mK),
the frequency of the NAMR can be in the range of GHz (e.g., $1$
GHz). If the qubit frequency $\omega_{0}$ and the Rabi frequency
$\Omega$ are taken as, e.g., $\omega_{0}=2$ GHz and $\Omega= 200$
MHz,  then the nonlinear parameter is $\kappa=8$ MHz. The
observation of the phonon blockade should be possible for a
quality factor $Q$ larger than $10^3$, which is in the NAMR
quality factor range $10^{3}\,\sim\,10^{6}$ of current
experiments. By engineering $\kappa$ as in
Refs.~\cite{Jacobs09,Gheri,Rebic}, $\kappa$ can be much larger
than $8$ MHz, then the phonon blockade should be easier to be
observed in our proposed system.

\section{Conclusions}

We have studied the quantum mechanics of the NAMR by coupling it
to a superconducting two-level system.  To demonstrate our
approach, a classical driving microwave is applied to the qubit so
that a dressed qubit is formed. If the Rabi frequency of the
driving field is strong enough, then the nonlinear phonon
interaction can be induced when the dressed qubit is in its ground
state. We mention that dressed charge qubits have been
experimentally realized~\cite{delsing1,delsing2}. The dressed
phase (see, e.g., Refs.~\cite{Wei04,martinis1,martinis2}) and flux
(see, e.g., Ref.~\cite{Liu05,ntt1}) qubits should also be
experimentally realizable.

The states of the nonlinear NAMR can be completely characterized
by the Cahill-Glauber $s$-parametrized quasiprobability
distribution (QPD). A state is considered to be nonclassical if it
is described by a $P$-function (QPD for $s=1$) that cannot be
interpreted as a probability density. As a drawback, the
$P$-function is usually too singular to be presented graphically.
Thus, other QPDs are often analyzed: If, for a given state, a QPD
with $s>-1$ is negative in some regions of phase-space, then the
state is nonclassical. We have shown that the Wigner function (QPD
for $s=0$) is always non-negative for nonclassical steady states
generated in our dissipative system. Thus, we have calculated the
$1/2$-parametrized QPD, being negative in some regions of
phase-space, which clearly indicates the nonclassical character of
the steady states generated in our NAMR system. Nevertheless, from
an experimental point of view, the quantum-state tomography of the
$1/2$-parametrized QPD is very challenging. Thus, we have proposed
another experimentally-feasible test of nonclassicality: the
phonon blockade.

We considered the case when the phonon self-interaction strength
$\kappa$ significantly exceeds the phonon decay rate $\gamma$. We
showed that when the NAMR oscillations are in the quantum regime,
the phonon transmission can be blockaded in analogy to the
single-photon blockade in a cavity~\cite{Imamoglu,Leonski94} or
Coulomb blockade for electrons~\cite{Coulomb}. When the phonon
blockade happens we also showed that a NAMR is in a nonclassical
state even if its Wigner function is non-negative. Therefore, the
nonclassicality of the NAMR can be demonstrated by the phonon
blockade, instead of trying to detect the tiny displacements when
the NAMR approaches the quantum limit. We further demonstrated that
the phonon blockade can be experimentally observed by measuring the
correlation spectrum of the electromotive force. All parameters in
our approach are within current experimental regimes and, therefore,
the quantum signature of the NAMR might be demonstrated in the near
future by using this proposed approach.

We have shown that the phonon blockade can be demonstrated by a
qubit-induced nonlinear NAMR. However, the temperature of the
environment, the decay rate of the NAMR, the driving current, and
the nonlinear coupling constant $\kappa$ limit the measured power
spectrum. To more efficiently observe the phonon blockade, the
following conditions should also be satisfied: (i) the temperature
should be low enough so that thermal excitations should be
negligibly small or the thermal energy is smaller than that of the
oscillating energy of the NAMR; (ii) the quality factor of the
NAMR should be high; (iii) the driving current through the NAMR
should be very weak, so that the heating effect induced by the
driving current can be neglected; (iv) the giant nonlinear
constant $\kappa$ of the NAMR might be more useful for the phonon
blockade, and this might be obtained using the approaches
explored, e.g., in Refs.~\cite{Jacobs09,Gheri,Rebic}. In our
proposal, the larger coupling constant $g$ between the qubit and
the NAMR corresponds to a larger $\kappa$, and the phonon blockade
should be more easily observable for larger $\kappa$. Also the
frequency of the NAMR should be large enough, so that the qubit
and the NAMR are in the large detuning regime, but the detuning
should not be extremely large.


\begin{acknowledgments}
FN acknowledges partial support from the Laboratory of Physical
Sciences, National Security Agency, Army Research Office, National
Science Foundation grant No. 0726909, JSPS-RFBR contract No.
09-02-92114, Grant-in-Aid for Scientific Research (S), MEXT
Kakenhi on Quantum Cybernetics, and Funding Program for Innovative
R\&D on S\&T (FIRST). YXL is supported by the National Natural
Science Foundation of China under Nos.~10975080 and 60836001. YBG
is supported by the NSFC Grant Nos.~10547101 and 10604002. CPS is
supported by the NSFC Grant No.~10935010. AM acknowledges support
from the Polish Ministry of Science and Higher Education under
Grant No.~2619/B/H03/2010/38. JB was supported by the Czech
Ministry of Education under Project No. MSM6198959213.
\end{acknowledgments}


\end{document}